\documentclass{article}

\pdfoutput=1

\usepackage{arxiv}

\usepackage[utf8]{inputenc} 
\usepackage[T1]{fontenc}    
\usepackage{hyperref}       
\usepackage{url}            
\usepackage{booktabs}       
\usepackage{amsfonts}       
\usepackage{nicefrac}       
\usepackage{microtype}      
\usepackage{lipsum}
\usepackage[numbers]{natbib}
\usepackage{bm}
\usepackage{amsmath}
\usepackage{graphics}

\title{Taking advantage of sampling designs in Bayesian spatial small area survey studies.}

\author{
    Carlos Vergara-Hernández   \\
    Vaccine research area \\
    FISABIO foundation \\
  Valencia (Spain)\\
  \texttt{\href{mailto:vergara_car@gva.es}{\nolinkurl{vergara\_car@gva.es}}} \\
   \And
    Marc Marí-Dell'Olmo   \\
    Environmental Quality and Intervention Service\\
    Agència de Salut Pública de Barcelona \\
  Barcelona (Spain)\\
  \texttt{\href{mailto:mmari@aspb.cat}{\nolinkurl{mmari@aspb.cat}}} \\
   \And
   Laura Oliveras \\
    Environmental Quality and Intervention Service\\
    Agència de Salut Pública de Barcelona \\
    Barcelona (Spain)\\
  \texttt{\href{mailto:lolivera@aspb.cat}{\nolinkurl{lolivera@aspb.cat}}} \\
   \And
    Miguel A. Martinez-Beneito   \\
    Department of Statistics and Operations Research. \\
    University of Valencia. \\
  Burjassot (Spain). \\
  and\\
  Joint research unit FISABIO-UV for the analysis of biomedical data\\
  \texttt{\href{mailto:miguel.a.martinez@uv.es}{\nolinkurl{miguel.a.martinez@uv.es}}} \\
  }

\begin{document}
\maketitle

\begin{abstract}
Spatial small area estimation models have become very popular in some contexts, such as disease mapping. Data in disease mapping studies are exhaustive, that is, the available data are supposed to be a complete register of all the observable events. In contrast, some other small area studies do not use exhaustive data, such as survey based studies, where a particular sampling design is typically followed and inferences are later extrapolated to the entire population. In this paper we propose a spatial model for small area survey studies, taking advantage of spatial dependence between units, which is the key assumption used for yielding reliable estimates in exhaustive data based studies. In addition, and in contrast to most spatial survey studies, we take the approach of also considering information on the sampling design and additional supplementary variables in order to yield small area estimates. This makes it possible to merge spatial and sampling based approaches into a common proposal.
\end{abstract}


\section{Introduction}

Spatial statistics has historically devoted considerable effort to deriving reasonable statistical estimates which take into account the geographic nature of the units of study. When sparse information is available, or simply when spatial dependence is strong, spatial models usually yield substantially improved estimates, which overcome many of the estimation problems that non-spatial small area methods entail. Unfortunately, spatial methods have usually focused on the modeling of outcome variables, as for example in the disease mapping literature \citep{Martinez-Beneito.BotellaRocamora2019}, rather than on covariates that could be used in subsequent regression models, which could explain the variability found in some spatial outcome variable. This limits the use of ecological regression models that could explain outcomes due to the scarcity of covariates with important potential explanatory capabilities. A clear example of this setting is lifestyle habits covariates in ecological regression problems, for which exhaustive population information rarely exists and these can be known at best throughout sampling-based health surveys. In this regard, taking advantage of spatial information in health surveys, or sampling-based surveys in general, becomes an important applied and methodological problem.

Frequently, survey studies follow particular sampling designs in order to achieve a desired precision for their estimates over the entire region of study or for some geographic/demographic groups. Nevertheless, if small area inference is pursued, these groups rarely coincide with those small areas since this would typically require huge sample sizes in order to achieve the desired precisions. This produces several problems when these surveys are used for inference at a small area level (smaller than that used for the sampling design), such as areas without any sampled unit or areas with unacceptably high variability on (direct) estimates of interest. In this context the use of spatial models is crucial.

In addition, when a sampling design has been followed, ignoring that design for inference seems a questionable choice since the available sample (and estimates) may thus no longer be representative of the overall population, over- or under-representing some population groups and thereby producing potentially biased or misleading results. As a solution, sampling theory proposes alternative estimates that take the followed sampling scheme into account. Therefore, taking the sampling design into account seems, in principle, as important as considering spatial dependence in small area studies; thus, model proposals which consider both features would be very welcome. However, regretfully, sampling designs in small area studies are typically planned for coarser spatial units than those considered, so most of the available sampling theory in the literature does not apply, at least in a direct manner, to small area studies.

Small area estimation literature has already dealt with sampling-based estimates for that geographical level. "This problem has been studied by many authors, albeit without much reference to the spatial nature of the geographical problem" (Lawson (2018) \cite{Lawson2018}, page 197), in addition to the evident benefits that spatial dependence brings in the small area context, as evidenced in the disease mapping literature. The small area estimation literature can be divided into two main approaches: design-based and model-based estimates. Design-based estimates rely on the inclusion probability of each individual in the sample, taking this factor into account by means of sampling weights (see for example Hansen and Hurwitz (1943)\cite{Hansen.Hurwitz1943} or Horvitz and Thompson (1952) \cite{Horvitz.Thompson1952}). In contrast, model-based estimates use some kind of model in order to take the sampling design into account \citep{Swensson.Saerndal.ea2003}. Nevertheless, spatial approaches to survey-based small area estimation usually follow hybrid approaches where (non-spatial) design-based estimates are derived, which take the sampling design followed into account, and spatial dependence is later induced on these estimates by means of a (spatial) smoothing model \citep{Chen.Wakefield.ea2014, Mercer.Lu.ea2019, Paige.Fuglstad.ea2020}. Although the uncertainty in the first stage of these models is usually considered for the second stage, the different processes modelled in both phases could interact; however, that interaction can hardly be corrected in these two-phase proposals. On the other hand, pure model-based estimates have also been proposed for survey-based small area estimation. From the Bayesian point of view, these models make use of hierarchical models in order to shrink group estimates that are based on a small number of statistical units \citep{Little1993, Park.Gelman.ea2004, Gelman2007}. Even though the flexibility of this approach makes it a very natural framework for inducing spatial dependence, at the same time that survey design issues could also be taken into account, while at the same time also taking survey design issues into account, spatial dependence has rarely been considered within this approach. In this paper we will follow the model-based approach, with the particular goal of including both survey design information and spatial dependence, at the same time, in our estimates.

Our general goal with this paper is to derive some sensible inference procedure for small area survey studies. As specific goals, we also intend to take spatial dependence and the sampling design into account in order to take advantage of these important features for our estimates. Furthermore, as a second specific goal, we also aim for such a procedure to allow inferences to be made at a small area level, smaller than that originally planned for the sampling design. Specifically, we will try to take advantage of the information used for the sampling plan and, if possible, all the related additional information that allowed us to yield enhanced estimators, including as much information as possible. In this paper we will focus on the specific, but frequently observed, case of estimating proportions of a binary variable in a collection of geographical units. Additionally, we will also assume that we will have some categorical auxiliary variables that will be helpful, or that simply should be considered by sampling design in order to improve those proportion estimates. The case of numerical output variables or covariates is left for future work, although the studied setting covers a high proportion of real settings in practice.

The paper in organized as follows. Section 2 introduces our approach in the trivial case of simple random sampling, where no particular sampling design is used to draw the survey sample. This section includes a real case study where several estimation proposals are compared. This will be the cornerstone for more complex sampling designs. Section 3 generalizes the simple framework introduced in Section 2 to more complex sampling designs. In particular, we show how to proceed in the very common setting of having a stratified random sample or if we wanted to use additional auxiliary variables for our inference. Section 3 also ends with a reanalysis of the case study analyzed in Section 2, incorporating the complexities considered in this new section. Finally, the paper concludes in Section 4 with a summary of some conclusions.

\section{Small area inference under simple random sampling}

Let us assume $I$ (small) spatial units of study, neighborhoods of a city from now on, for example, in accordance with the following case studies. Let $Y \in \{0,1\}$ be a random variable, whose mean (proportion of 1s in the corresponding population) we would be interested to know for each of the $I$ mentioned spatial units. Let us also assume that $Y$ has been observed for just a sample of individuals for each spatial unit, sampled under simple random sampling (SRS) for all these neighborhoods. We will denote the corresponding sample size for each neighborhood as $n_i\;(i=1,\ldots,I)$ of a total of $N_i\;(i=1,\ldots,I)$ individuals living in each of those neighborhoods. In that case, under a sampling theory perspective, we would like to know, or estimate from our sample, the proportions $\pi_i=E(Y_i)=N_i^{-1}\sum_{j=1}^{N_i} Y_{ij}$ where $Y_{ij}$ stands for the observed value of $Y$ for the $j$-th individual of neighborhood $i$.

Under SRS, those proportions may be estimated as simply the sample mean $\hat{\pi}_i^{SRS}=\bar{Y}_{i\cdot}=(n_i)^{-1}\sum_{j\in\partial_i}Y_{ij}$ of $Y$ at each neighborhood, where $\partial_{i}$ indexes the set of individuals of neighborhood $i$ which have been sampled. This elemental result can be found, for example, on page 35 of Lohr (2010) \cite{Lohr2010}, among many other sampling theory books. Nevertheless, we will reference sampling theory results according to Lohr's book (adding the corresponding book page) to be more precise about our citations. The sample mean in that scenario can be shown to be an unbiased estimator of each $\pi_i$ (see page 52 of Lohr (2010) \cite{Lohr2010}), with variance:
\begin{equation}\widehat{Var}(\hat{\pi}_i^{SRS})=\frac{\hat{\pi}_i^{SRS} (1-\hat{\pi}_i^{SRS})}{n_i}(1-\frac{n_i}{N_i}),\label{EqVars}\end{equation}
where the second term in this product takes into account the finite population nature of each neighborhood (page 36 of Lohr (2010) \cite{Lohr2010}). Additionally, motivated by the hypothetical Normality of sample means for large sample sizes, a $100(1-\alpha)\%$ confidence interval for $\pi_i$ is frequently derived as simply $\hat{\pi}_i^{SRS}\pm t_{1-\alpha/2}\sqrt{\widehat{Var}(\hat{\pi}_i^{SRS})}$, where $t_{1-\alpha/2}$ is the $1-\alpha/2$ quantile of a $t$ distribution with $n_i-1$ degrees of freedom (page 43 of Lohr (2010) \cite{Lohr2010}).

The $\hat{\bm{\pi}}^{SRS}=(\hat{\pi}_1^{SRS},\ldots,\hat{\pi}_I^{SRS})$ estimator may also be justified from other points of view beyond sampling theory. For example, it can be easily shown that $\hat{\pi}_i^{SRS}$ is also the maximum likelihood estimator (MLE) of $\pi_i$ for the frequentist model:
$$Y_{ij}\sim Bernoulli(\pi_i),\;i=1,\ldots,I;\;j\in\partial_i,$$
and is also the MLE for the abridged version of this model:
$$O_i\sim Binomial(\pi_i,n_i),\;i=1,\ldots,I,$$
where $O_i=\sum_{j=1}^{n_i}Y_{ij}$. In addition, under a Bayesian perspective, $\hat{\bm{\pi}}^{SRS}$ may be also viewed as the posterior mode of the models with either of the two previous data likelihoods and with $\pi_i,\;i=1,\ldots,I,$ independently distributed with $Uniform(0,1)$ prior distributions. This is a consequence of the equivalence of the frequentist MLE and the Bayesian posterior mode when independent uniform prior distributions are used for the components of $\bm{\pi}$. These results further support the convenience and use of $\hat{\bm{\pi}}^{SRS}$ as an (unbiased) estimator of $\bm{\pi}$ under SRS.

According to all this, $\hat{\bm{\pi}}^{SRS}$ may be considered an unquestionable estimator of $\bm{\pi}$ under SRS. Nevertheless, this estimate may sometimes not be so well suited. Specifically, when the spatial units of study are small, $\hat{\bm{\pi}}^{SRS}$ may become unreliable (excesive variance) and yield inappropriate estimates. This is something similar to what happens in disease mapping studies where the most common risk estimator, Standardized Mortality Ratio (SMR), given by $O_i/E_i$ where $O_i$ and $E_i$ are the corresponding observed and expected deaths for unit $i$, is discouraged for small area studies. SMRs also coincide with the MLE of $\bm{\lambda}$ for the model:
$$O_i\sim Poisson(\lambda_i E_i),\;i=1,\ldots,I,$$
or the mode of the Bayesian model assuming that same data likelihood and $\lambda_i\sim Uniform(0,\infty),\;i=1,\ldots,I$ (page 111 of \cite{Martinez-Beneito.BotellaRocamora2019}). This illustrates the parallelism between the small area estimation problems of $\hat{\bm{\pi}}^{SRS}$ and those of SMRs in their different contexts.

A large body of literature has been published in the disease mapping area that tries to fix the small area estimation problems described. The main proposals to overcome the problems inherent in that approach come from Bayesian hierarchical models, where the independent uniform prior distributions for the components of $\bm{\lambda}$ are susbtitued by a spatial prior distribution. That distribution shares information between neighboring units and therefore uses neighbors' information for each area as a secondary information source, therebore yielding more reliable risk estimates. Different conditional auto-regressive (CAR) distributions have been proposed for this purpose, such as the intrinsic CAR \citep{Kunsch1987} or lCAR distributions \citep{Leroux.Lei.ea1999}, or even the combination of spatial and non-spatial processes \citep{Besag.York.ea1991}.

Something similar could be proposed to estimate $\bm{\pi}$ in survey sampling contexts, as introduced above, to improve the $\hat{\bm{\pi}}^{SRS}$ traditional estimates. Being more precise, we could estimate the proportions $\bm{\pi}$ as the estimates arising from the following Bayesian hierarchical model:

\begin{eqnarray*}
  Y_{ij} &\sim & Bernoulli(\pi^*_i),\;i=1,\ldots,I;\;j\in\partial_i \\
  logit(\pi^*_i) &=& \mu+\theta_i \\
  &\ldots & \\
\end{eqnarray*}
where $\bm{\theta}=(\theta_1,\ldots,\theta_I)$ is a vector of spatially structured random effects following, for example, one of the CAR-based spatial processes mentioned. For this spatial model the population proportions $\pi_i$ could be estimated as:
\begin{equation}\hat{\pi}^{Spatial}_i=\frac{n_i}{N_i}(\bar{Y}_{sampled})_i+\frac{N_i-n_i}{N_i}(\bar{Y}_{unsampled})_i,\label{EqFinite}\end{equation}
where $(\bar{Y}_{sampled})_i$ and $(\bar{Y}_{unsampled})_i$ correspond, respectively, to the mean for the sampled and unsampled populations for the $i$-th spatial unit. The first term on the right hand side of this expression is known for each neighborhood and the second term can be sampled from the corresponding predictive distribution of the model for each step of the MCMC process. This estimator takes into account the finite sample feature of each proportion since when the sample size grows, the first component of the previous expression, which has no uncertainty, increases its weight, reducing therefore the variability of $\hat{\pi}^{Spatial}_i$. In any event, if the sampled population fraction ($n_i/N_i$) is small, $\hat{\pi}^{Spatial}_i$ will be very similar to the model parameters $\pi^*_i$, so these values could alternatively be used as $\pi_i$ estimates for simplicity.

In principle, this model could fix the small area estimation problems that $\hat{\bm{\pi}}^{SRS}$ display, similarly to how SMRs estimates are typically improved in disease mapping studies. Unfortunately, in contrast to $\hat{\bm{\pi}}^{SRS}$, the proportion estimates arising from this model will no longer be unbiased, since these proportion estimates will be shrunk toward those of their neighbors. Nevertheless, the more efficient use of information in spatial models, which is shared between nearby units, should yield more accurate proportion estimates than raw sample means, which will compensate for the bias introduced by these estimators.

\subsection{Estimating the proportion of low- and middle-income countries population for Barcelona neighborhoods: A real case study}

In this case study, which will be revisited afterwards, we are going to compare different estimates assuming SRS, or simply ignoring the sampling design, to the estimation of the proportion of people from low- and middle-income countries (LMIC for short from now on) for neighborhoods in the city of Barcelona, Spain. Thus, the outcome variable $Y$ for this study is equal to 1 if the surveyed person was born in a LMIC, and 0 otherwise. Barcelona is divided into 73 neighborhoods with populations varying from 631 to 58,054 inhabitants, with a median of 20,533 people. The survey data used for this case study corresponds to the 2016 Barcelona Health Survey, with a sample size of around 4000 people. This survey was designed to achieve a given precision level for each of the 10 districts of the city, so our units of study are smaller than those corresponding to the originally planned design. The sample size for the different neighborhoods of the city vary between 1 and 226, with a median value of 46, yielding a surveyed population of 2.4 people for each 1,000 city inhabitants. Interestingly, also for 2016, the Barcelona register of inhabitants, which collects population information for the entire city (not just a sample) contains information on the birth country of the city inhabitants. Therefore, we could alternatively obtain from this source the proportion of LMIC people for each neighborhood throughout this exhaustive sample, which could be used as a gold standard (the *gold standard* $\bm{\pi}$ from now on) for assessing the quality of the different survey estimates. All the details (Rmarkdown document with the \texttt{R} code run) of the following and subsequent analyses in this paper can be found in the supplementary material of the paper (\texttt{\href{https://www.uv.es/mamtnez/Sampling.html}{\nolinkurl{https://www.uv.es/mamtnez/Sampling.html}}}).

By using the survey data, we have derived four different estimates of the proportion of LMIC population at the neighborhood level $\bm{\pi}=(\pi_1,\ldots,\pi_I)$. First, we have derived a \textit{frequentist SRS} estimate of those proportions, $\hat{\pi}^{SRS}$ as defined above, which is simply the sample mean of $\{Y_{ij}:\;j\in\partial_i\}$. Variance estimates ($\widehat{Var}(\hat{\bm{\pi}}^{SRS})$) and 95
$$\frac{(O_i+1)-1}{(O_i+1)+(n_i-O_i+1)-2}=\frac{O_i}{n_i},$$
which could be used once again as proportion estimates for this model. This Bayesian point estimate coincides with $\hat{\pi}^{SRS}_i$. Nevertheless, for the Bayesian case, credible intervals for $\pi_i$ may be derived from those posterior distributions, which would avoid the Normality assumption of the frequentist estimators not so suited, possibly, for small areas. These credible intervals would take into account the Binomial nature of the outcome variable, which could yield, in principle, some advantages. Finally, we have derived two spatial proportion estimates corresponding to the Bayesian spatial model introduced previously, which considers those proportions as dependent quantities in contrast to the other two alternatives. Specifically, we consider the spatial estimate accounting for the finite population character of the sample ($\hat{\bm{\pi}}^{Spatial}$ in expression \ref{EqFinite}) and, as an alternative, the proportion estimates which ignore that character ($\bm{\theta}^*$). As point estimates we use the posterior distribution of those variables. An lCAR distribution, as introduced by Leroux et al. (1999) \cite{Leroux.Lei.ea1999}, is used to induce spatial dependence (vector $\bm{\theta}$) in the proportion estimates for this model.

Table \ref{tableSRS} shows several statistics for these proportion estimates. To assess their accuracy we have calculated the correlation with the gold standard, drawn from the Barcelona register of inhabitants, and (the square root of) the mean squared error for each proportion estimate. Furthermore, we have assessed the precision of those estimates by means of their 95\% confidence/credible interval (CI) lengths. Additionally, we have also assessed the empirical coverage of those CIs as the proportion of those 73 intervals that contained the corresponding gold standard proportion $\pi_i$, which we would expect to be around 0.95. We have also measured the spatial dependence of the estimates by means of the Moran's $I$ statistic \citep{Moran1948}, which should be compared with the gold standard's ($\bm{\pi}$) Moran's I, equal to 0.202. Finally, we have assessed the divergence of the Moran's $I$ statistics for these estimates with that of the gold standard $\bm{\pi}$. We have done this by evaluating the posterior probability of the Moran's $I$ of these statistics being lower than the Moran's $I$ of $\bm{\pi}$. For the Bayesian cases, this has been done by drawing  samples of the proportions posterior distributions and then calculating the Moran's $I$ posterior distribution from these. For the frequentist case, we have repeatedly sampled independent values of a t distribution with $n_i-1$ degrees of freedom of mean $\hat{\pi}_i^{SRS}$ and variance $\widehat{Var}(\hat{\pi}_i^{SRS})$, the distribution implicitly assumed for deriving CIs. For these samples we have calculated the corresponding Moran's $I$, which have been used to estimate the probability of that statistic being lower than the Moran's $I$ of $\bm{\pi}$.

\begin{table}[h]
\centering
\caption{Assessment criteria for each proportion estimate assuming simple random sampling}
\begin{tabular}{ |c|c|c|c|c| }
 \hline
 Statistic & $\hat{\bm{\pi}}^{SRS}$ & \textit{Bayes SRS} & $\hat{\bm{\pi}}^{Spatial}$ & $\hat{\bm{\pi}}^{*}$\\ \hline
 $Cor(\bm{\pi},\hat{\bm{\pi}})$ & 0.612 & 0.612 & 0.747 & 0.746 \\
 $\sqrt{MSE}$ & 0.083 & 0.083 & 0.049 & 0.049\\
 95\% CI length & 0.271 & 0.252 & 0.172 & 0.172\\
 95\% CI Coverage & 0.808 & 0.959 & 0.945 & 0.945\\
 Moran's I & 0.129 & 0.129 & 0.233 & 0.235\\
 P(estimated I<real I) & 0.986 & 0.969 & 0.671 & 0.663\\
 \hline
\end{tabular}
\label{tableSRS}
\end{table}

The first two lines of Table \ref{tableSRS} show how the point estimates (posterior means) of the spatial model ($\hat{\bm{\pi}}^{Spatial}$ and $\hat{\bm{\pi}}^{*}$) outperform those of the independent SRS estimates in terms of both correlation and MSE with respect to $\bm{\pi}$. This illustrates how sharing information between spatial units improves point estimates when dealing with small areas and how that increase in the information used outperforms the potential bias of those estimates. The benefits of spatial estimates are also evident in the CI lengths, as the CIs for this model are substantially narrower than for the other two alternatives. Despite the enhanced precision of the spatial estimates, their 95\% CI coverage is substantially better than that of the frequentist SRS, which is lower than it ought to be. Interestingly, the \textit{Bayes SRS} estimate already shows an evident improvement in the CI coverage, similar to that of the spatial estimates. We could conclude therefore that the coverage problems of the frequentist SRS estimates come from the Normality assumption, which, for low/moderate sample sizes, such as those typically arising from small areas, does not hold in practice. Finally, the last two lines of Table \ref{tableSRS} summarize the spatial dependence of the proportion estimates. As shown, the spatial model shows a clear increase in the spatial dependence of the estimates, with its Moran's $I$ being nearly twice that of its alternatives and much closer to the gold standard Moran's $I$ (0.202). Moreover, for the spatial estimates the gold standard Moran's $I$ falls around the 67\% percentile of the posterior distribution, which highlights the compatibility of the spatial dependence of those estimates with that of $\bm{\pi}$. In contrast, the gold standard Moran's $I$ falls around the 97\% percentile of the estimates for the other two non-spatial alternatives, suggesting that these estimates display weaker spatial dependence than they ought to. Finally, Table \ref{tableSRS} points out the equivalence in practice of the finite population $(\hat{\bm{\pi}}^{Spatial})$ and general $(\hat{\bm{\pi}}^{*})$ spatial estimates, which suggests that for the sampling fraction in this study (2.4 sampled individuals for each 1,000 inhabitants) no finite population correction is really needed.

\section{Small area inference for more complex sampling designs}

In the following, we will consider the inclusion of auxiliary variable(s) information in survey based inference, regardless of whether or not this information is used for designing the sample of the study (stratified sampling).

\subsection{Small area inference for stratified sampling designs}

Despite the simplicity of SRS, which is the simplest scheme followed for collecting survey samples, this is not such a common sampling scheme in practice. Instead, more complex designs are usually followed for these studies. Stratified sampling, for example, is a much more frequent choice, where samples are divided into several population groups and their estimates are then merged in order to derive compound indicators with a given precision. This process could be done for either the whole population or particular population groups, such as specific geographical areas. Unfortunately, when a sample is planned, it will rarely be planned for a small area division of the region of study but rather for a coarser spatial division, or even the whole region, since otherwise the required sample size will typically be unattainable. Our goal will now be to use those stratified samples to make inferences at a higher spatial disaggregation level than that originally planned (small areas), taking into account the specific stratified sampling scheme followed.

The main goal of stratification is to reduce the variance of the estimates by combining the information of several homogeneous groups that, in principle, should have low variance. From this point on, we will consider that a stratified sampling design has been used to gather a survey sample and a SRS scheme has been used in order to collect the corresponding sample within each stratum. More specifically, and without loss of generality, we will assume that the sampling strata correspond to age groups. This framework generalizes the previous SRS scheme to a much more common and general setting in practice. Additionally, from now on we assume that the stratified sample was originally collected for the whole region of study in order to estimate one proportion for that whole region with a given precision, but we want to now use that sample to estimate that same proportion, as accurately as possible, for the units of a small area division of the region of study. If an intermediate spatial division, coarser than the small areas of interest, was considered for the sampling design, all the following would remain valid; nevertheless, for clarity we will focus on the simplest case of a single stratified sample for the whole population. That is, the following is valid regardless of the spatial division considered, if any, to design the sample of the study.

Sampling theory states (see page 73 of Lohr (2010) \cite{Lohr2010}) that under stratified sampling, with SRS within each age group $k(=1,\ldots,K)$, an unbiased estimate of $\pi=E(Y)=N^{-1}\sum_{j=1}^N Y_j$, the overall proportion of $Y=1$ over the whole population, could be derived as:
\begin{equation}\hat{\pi}^{Str}=N^{-1}\left(\sum_{k=1}^K \bar{Y}^{Str=k} N^{Str=k} \right),\label{EqStr}\end{equation}
where $\bar{Y}^{Str=k}$is the sample mean of $Y$ on age group $k$ and $N^{Str=k}$ the population size of that same age group, with $N=N^{Str=1}+\ldots+N^{Str=K}$. Note that each of the terms in the sum on the right-hand-side of this expression $(N^{Str=k} \bar{Y}^{Str=k})$ can be seen as a ratio estimator (See Chapter 4 of Lohr (2010) \cite{Lohr2010}) for the total of $Y$ for each stratum, where that ratio ($\bar{Y}^{Str=k}$) compares the total people with $Y=1$ against the total sampled people in that stratum. Afterwards, this ratio is combined with the auxiliary design information (population size $N^{Str=k}$) in order to derive a ratio estimate (page 119 of Lohr2010) of the total of people with $Y=1$ in the corresponding age group. Therefore, the stratified estimator (\ref{EqStr}) may be considered as a particular case of the ratio estimator, which will soon be important below.

Unfortunately, stratified sampling on small areas poses additional estimation problems. On the one hand, stratified sampling collects several samples (strata) for each unit of analysis, instead of a single sample, so this increases the sample size requirements to derive estimates of this kind. Furthermore, when a stratified sampling scheme is used to derive areal estimates, the spatial units originally considered for the sampling design (let us say US states) are frequently larger than the small areas of interest (let us say counties), since otherwise the required sample size to achieve the desired precision would not be attainable. Having a sampling design at a coarser spatial level than the small areas of interest, could easily lead to some o the age groups not being sampled in some of those small areas, which would make the direct estimation of the parameter of interest $\hat{\pi}^{Str}$ unfeasible for those areas. Additionally, when small area inference is undertaken in that setting, the original stratification weights, which were initially planned for larger units, may no longer be valid for deriving estimates for those small areas, so the available sample could seem useless for maiking inferences about those small areas. Our goal will now be to solve these problems when small area inference is pursued on a stratified sample designed for a broader (if any) geographical division.

In that case, let us denote $N_{i}^{Str=k}$ and $n_i^{Str=k}$ the total and sampled populations, respectively, for the $i$-th neighborhood and $k$-th age group, $i=1,\ldots,I;\;k=1,\ldots,K$. Note that $N_{i}^{Str=k}$ will generally be known in advance of the sample selection, since the sampling design is organized according to these values, but not $n_i^{Str=k}$, which will depend on the sample drawn. Let us also denote $Y_{ij}^{Str=k};\;i=1,\ldots,I;\;j=1,\ldots,n^{Str=k}_i;\;k=1,\ldots,K,$ the observed value for the binary variable $Y$ of interest, for the $j$-th individual of the $k$-th age group on the $i$-th neighborhood. Let us also assume that our goal will now be to estimate $\pi_i=E(Y_i)=N_i^{-1}\sum_{k=1}^{K}(\sum_{j=1}^{N^{Str=k}_i}Y^{Str=k}_{ij})$, which is equivalent to $N_i^{-1}\sum_{j=1}^{N_i}Y_{ij}$ if the stratified design/notation is ignored.

Expression (\ref{EqStr}) establishes how to use information on the stratification variable to compound estimates for the whole population. Unfortunately, that expression would in principle be useless for deriving areal estimates for a small area division of the region of study which does not coincide with that used for the sampling design. Nevertheless, we mentioned that Expression (\ref{EqStr}) can be viewed simply as a combination of ratio estimators, beyond its stratified nature, and that combination could now be used now as a post-stratification procedure to derive stratified small area estimators. Thus, let $\bar{Y}_i^{Str=k}=(n_i^{Str=k})^{-1}\sum_{j=1}^{n_i^{Str=k}}Y_{ij}^{Str=k}$ be the sample mean for stratum $k$ and the (small) spatial unit $i$. That estimate can be also expressed as: $(\sum_{j=1}^{N_i^{Str=k}}X_{ij}^{Str=k}Y_{ij}^{Str=k})/(\sum_{j=1}^{N_i^{Str=k}}X_{ij}^{Str=k})$
, where $X_{ij}^{Str=k}=1$ if individual $j$ of stratum $k$ and spatial unit $i$ is included in the sample, and 0 otherwise. For this last expression the denominator is also stochastic, since SRS is used over an upper geographical level and we do not know how many of the elements of the $k$-th age group in the sample will correspond to spatial unit $i$. Therefore, $\bar{Y}_i^{Str=k}N_i^{Str=k}$ would simply be a ratio estimator, supported by sampling theory, of the $Y$ total for age group $k$. Therefore, the sum of these ratio estimators over all the age groups $\sum_{k=1}^K \bar{Y}_i^{Str=k}N_i^{Str=k}$ could be used to estimate the total of $Y$ over the entire spatial unit $i$, and:
\begin{equation}\hat{\pi}^{Str}_i=N^{-1}\sum_{k=1}^K \bar{Y}_i^{Str=k}N_i^{Str=k},\label{EqStrOther}\end{equation}
could be used as a combined ratio estimator, or stratified estimator, of $\pi_i=E(Y_i)$ suitable for a different (small area) spatial division than that considered for the stratified sampling design. This makes the available sample, originally planned for a coarser spatial division, also useful for a small area division of the region of study.

Although Expression (\ref{EqStrOther}) allows $\pi_i$ to be estimated for any division of the region of the study, taking the stratified nature of the sample into account, it does not solve the small area estimation problems that the size of the units of study could pose. These problems should be similar to, or possibly worse than, those illustrated in Section 2. If independence was assumed at the inference stage for each age group and spatial unit, then $\bar{Y}_i^{Str=k}$ in Expression (\ref{EqStrOther}) may sometimes not be computed if no sampled units were available for that particular age group and spatial unit combination. Additionally, if the number of sampled units for that combination was simply small, although greater than 0, which could occur frequently in practice, this would make $\bar{Y}_i^{Str=k}$ inaccurate, and that variability would also be transferred to $\hat{\pi}^{str}_i$. Thus, small area modeling becomes much more necessary in stratified sampling schemes than for SRS survey samples. In SRS, small area problems were alleviated by the modeling of the vector $\bm{\theta}$, considering their components as spatially dependent variables. Now we have a matrix $\mathbf{E(Y)}=(E(Y_i^{Str=k}))_{i,k=1}^{I,K}$ to estimate, instead of a single vector, which we could model in several different manners, possibly considering spatial dependence among its cells.

Being more precise, let us denote the outcome variable for the $j$-th sampled individual of spatial unit $i$ and age group $k$ as $Y_{ij}^{Str=k}$. We could then model $Y_{ij}^{Str=k}$ as a Bernoulli variable of probability $(\pi^*)_{i}^k$, where that probability could be modeled, for example, as:
$$logit((\pi^*)_{i}^{Str=k})=\mu+\psi_{i}+\phi_{k}.$$
In this model proposal, age ($\bm{\phi}$) and spatial ($\bm{\psi}$) effects could be modeled as usual, thus $\phi_1$ could be fixed to 0 while an improper uniform prior could be considered for $\phi_2,\ldots,\phi_K$ and a spatial CAR pior, such as the Leroux et al.'s lCAR, could be used to induce spatial dependence on $\bm{\psi}$. Once this model has been fitted, we could estimate the probabilities $\hat{\pi}_i^{Str=k}$ for each neighborhood and stratum, either taking a finite sample correction into account, or not, like for SRS. Therefore, the overall probability for each neighborhood could be defined as:
\begin{equation}\hat{\pi}^{StrSmall}_i=N^{-1}\sum_{k=1}^K \hat{\pi}_{i}^{Str=k}N_i^{Str=k},\end{equation}
which sums the estimated totals for every stratum. We could use $\hat{\pi}^{StrSmall}_i$ as an alternative estimate to $\hat{\pi}^{Str}_i$, which, in principle, could solve the small area estimation problems that the direct stratified estimator could exhibit.

Note that the modeling of $logit((\pi^*)_{i}^{Str=k})$ proposed above is just a simple proposal supposedly solving the original small area estimation problems of $\hat{\pi}^{Str}_i$. Clearly, that model induces strong assumptions about $\bm{\pi}$ that could alternatively be relaxed by considering for example a multivariate CAR process \citep{MacNab.2018} for the age groups used to stratify the sample. This would be less restrictive than the proposal above as it would allow age-neighborhood interactions to be reproduced if these were required. Nevertheless, the above proposal would in principle be more parsimonious than estimating a different spatial pattern per age group; therefore, for this reason and to save space, we will limit this work to exploring the benefits of the additive effects proposal of age and neighborhood mentioned above. However, if age-neighborhood interactions were suspected, an alternative multivariate CAR modeling of the age and neighborhood effects could be preferred over our proposal.

It is also worth mentioning that an abridged version of the model above could alternatively be proposed, as already mentioned for the SRS case, with important computational benefits if computation was an issue due to the dimensions of the data set of study. Thus, the Bernoulli likelihood of the model above could be substituted by a Binomial distribution on the number of observed people $O_i^{Str=k}$ with $Y_{ij}^{Str=k}=1$ (i.e. $O_i^{Str=k}=\sum_{j=1}^{n^{Str=k}_i}Y_{ij}^{Str=k}$) for the corresponding age group and neighborhood as follows:
$$O_i^{Str=k} \sim Binomial((\pi^*)_i^{Str=k},n_i^{Str=k}),\;i=1,\ldots,I,\; k=1,\ldots,K$$
and once again:
$$logit((\pi^*)_i^{Str=k})=\mu+\psi_i+\phi_k.$$
In this case, $\bm{\phi}$ and $\bm{\psi}$ could be modeled in the same manner as for the previous model. Both models can be shown to have the same data likelihood and thus their equivalence. Although this abridged model could be more convenient from a computational point of view, the consideration of additional auxiliary variables (as for the example below) would mean a major change in the formulation of this model since $O_i^{Str=k}$ and $n_i^{Str=k}$ would need to be recomputed for all the groups of the auxiliary variables. In contrast, considering an additional auxiliary variable for the original Bernoulli implementation would pose no particular problems in terms of the model implementation, so in that case the original model could be more convenient in practice. In fact, that model will be used for the case study developed below. In any event, both formulations should be borne in mind and used depending on their appropiateness for each particular problem.

\subsection{Small area inference with auxiliary variables}

In addition to stratified sampling, a similar proposal to that above could be used to take into account the information of additional auxiliary variables, even though these were not originally used to design the sample. In fact, as mentioned, stratification is just a particular case of the use of auxiliary variables, specifically in the sampling design phase of the survey. We have seen that poststratification has been the key to deriving stratified estimators for units of study that were not originally considered in the survey design. This same idea can also be used to include auxiliary variables in the inference, it also being possible to consider spatial dependence in that case.

Let us now assume that our survey collected, in addition to our variable of interest $Y$, a second categorical variable $Z\in\{1,\ldots,K\}$ for each individual in the sample. Let us also assume that we knew, from a second information source, the population $N_i^{Z=k}$ belonging to each of the $K$ categories of $Z$ for each of the small areas $i$ that make up the region of study, i.e. $N_i^{Z=k}=\sum_{j=1}^{N_i}1_{Z_{ij}=k};\;i=1,\ldots,I;\;k=1,\ldots,K$, where $1_{Z_{ij}=k}$ is equal to 1 if $Z=k$ for individual $j$ of spatial unit $i$, and 0 otherwise. In that case, similarly to the stratified setting, we could use a ratio estimator to estimate the population with $Y=1$ for each spatial unit and group of the auxiliary variables as follows:
$$(\hat{\tau}_Y)_i^{Z=k}=\left(\frac{\sum_{j=1}^{n_i}(Y_{ij}^k\cdot 1_{Z_{ij}=k})}{\sum_{j=1}^{n_i}1_{Z_{ij}=k}}\right)N_i^{Z=k}= (\bar{Y}|Z_{ij}=k)N_i^{Z=k}.$$
In addition, we could also estimate $\pi_i=N_i^{-1}(\tau_Y)_i$, the percentage of people with $Y=1$ in spatial unit $i$ as:
\begin{equation}\hat{\pi}^{Ratio}_i=N_i^{-1}(\hat{\tau}_Y)_i=N_i^{-1}\left(\sum_{k=1}^K (\hat{\tau}_Y)_i^{Z=k}\right)=N_i^{-1}\sum_{k=1}^K (\bar{Y}|Z_{ij}=k)N_i^{Z=k}.\label{EqRatio}\end{equation}
This estimator would take advantage of the information on the total population for each value of $Z$, $N_i^{Z=k}$, that we know for each spatial unit. Once again, this estimator allows us to include auxiliary variables information on our estimates in a similar manner to how Expression (\ref{EqStrOther}) did it with the stratification variable. Nevertheless, this new estimate would pose the same small area estimation problems as those posed by the estimator in Expression (\ref{EqStrOther}). Luckily, they could be solved in exactly the same way as done there, by means of small area hierarchical modeling that considers spatial dependence. Thus, the hierarchical model proposed to take advantage of stratified sampling would one again be valid also for the poststratification of auxiliary variables by simply changing the $N_i^{Str=k}$ used in Expression (\ref{EqStrOther}) to the totals $N_i^{Z=k}$ of each group for the auxiliary variable.

\subsection{Small area inference with stratification and  auxiliary variables}

We have just seen that accounting for stratified designs in spatial divisions unplanned during the sampling design and using auxiliary variables, also ignored in the sampling design, does not pose any difference in terms of their  statistical analysis. We are now going to consider how to combine these two tools in order to obtain enhanced estimators of use in many real settings. We will consider the case of having two auxiliary variables, regardless of whether or not they are used to design the sample (stratification variables), since the case of considering more than two variables would be closely similar.

Let us consider that we have two auxiliary variables, $Z_1$ with values in $\{1,\ldots,K_1\}$ and $Z_2$ with values in $\{1,\ldots,K_2\}$. In that case, our quantities of interest will be:
\begin{align}
\pi_i&=N_i^{-1}\sum_{j=1}^{N_i} Y_{ij}=N_i^{-1}\sum_{j=1}^{N_i}\sum_{k_1=1}^{K_1}\sum_{k_2=1}^{K_2} Y_{ij}1_{(Z_1)_{ij}=k_1}1_{(Z_2)_{ij}=k_2}=\nonumber\\
&= N_i^{-1}\sum_{k_1=1}^{K_1}\sum_{k_2=1}^{K_2}\pi_i^{Z_1=k_1,Z_2=k_2}N_i^{Z_1=k_1,Z_2=k_2},\label{FullEstimate}
\end{align}
where $\pi_i^{Z_1=k_1,Z_2=k_2}=(N_i^{Z_1=k_1,Z_2=k_2})^{-1}\sum_{j=1}^{N_i}Y_{ij}1_{(Z_1)_{ij}=k_1}1_{(Z_2)_{ij}=k_2}$ stands for the proportion of times that $Y=1$ in neighborhood $i$ when $Z_1=k_1$ and $Z_2=k_2$ and $N_i^{Z_1=k_1,Z_2=k_2}$ is the population corresponding to those values of the auxiliary variables. As introduced in the case of a single stratification or auxiliary variable, $\pi_i^{Z_1=k_1,Z_2=k_2}$ for $k_1=1,\ldots,K_1$ and $k_2=1,\ldots,K_2$ could be estimated by the corresponding means $\bar{Y}_i^{Z_1=k_1,Z_2=k_2}$ in the available sample. In that case, Expression (\ref{FullEstimate}) would produce an estimate $\hat{\pi}_i$ that would take into account both auxiliary variables. Nevertheless, that choice would show clear small area problems, similarly to the same problems mentioned above. As a consequence, $\pi_i^{Z_1=k_1,Z_2=k_2}$ could be estimated instead by means of a spatial hierarchical model that would solve the small area problems that these indicators could exhibit.

The problem in practice with the estimator in Expression (\ref{FullEstimate}) is that $N_i^{Z_1=k_1,Z_2=k_2}$ will not be so frequently available in population databases when two or more auxiliary variables are combined. In that case, if the distribution of $Z_1$ and $Z_2$ in the population are reasonably independent, that is:
$$N_i^{Z_1=k_1,Z_2=k_2}\approx \frac{N_i^{Z_1=k_1}\cdot N_i^{Z_2=k_2}}{N_i},$$
then that estimator could be restated as:
\begin{equation}N_i^{-1}\sum_{k_1=1}^{K_1}\sum_{k_2=1}^{K_2}\pi_i^{Z_1=k_1,Z_2=k_2}\frac{N_i^{Z_1=k_1}\cdot N_i^{Z_2=k_2}}{N_i}=
N_i^{-1}\sum_{k_1=1}^{K_1}\left(N_i^{-1}\sum_{k_2=1}^{K_2}\pi_i^{Z_1=k_1,Z_2=k_2}N_i^{Z_2=k_2}\right)\cdot N_i^{Z_1=k_1}.\label{IndEstimate}\end{equation}
The term within the parenthesis in this expression reproduces, for each value of $Z_1$, a ratio estimator for $Z_2$ conditioned to that particular value of $Z_1$. Similarly, the rest of that expression provides a second ratio estimator, in this case for $Z_1$, which combines, according to this variable, the conditional ratio estimators previously calculated for $Z_2$. Therefore, this estimator may be viewed as a sequential ratio estimator of each of the auxiliary variables considered, where the information on each of the auxiliary variables is progressively included into the full estimator. Clearly, the order of the sums in Expression (\ref{IndEstimate}) could be reversed, thus the order used to include the auxiliary variables into the full estimator does not have any impact on it.

Both estimators in Expressions (\ref{FullEstimate}) and (\ref{IndEstimate}) are two separate alternatives, although the suitability of the second depends on the adequacy of the independence assumption for $Z_1$ and $Z_2$. Using one or the other in practice will depend on the availability of population information for the combination of both auxiliary variables, $N_i^{Z_1=k_1,Z_2=k_2};\; i=1,\ldots,I;\;k_1=1,\ldots,K_1;\;k_2=1,\ldots,K_2$ and/or the validity of the independence hypothesis between $Z_1$ and $Z_2$, if this may be assessed, otherwise it will just have just to be assumed if population information on their combination is not available.

\subsection{Estimating the proportion of LMIC people for Barcelona neighborhoods: A reanalysis}

We are going to resume now the study of the estimation of LMIC population for Barcelona neighborhoods, already examined in the previous section. The survey sampling used in the Barcelona health survey originally followed a stratified design, with strata corresponding to five age groups ($\{0-19, 20-39, 40-59, 60-79, 80-\ldots\}$) and sex, which was ignored in the previous case study. Additionally, the total sample size was divided into 10 groups of around 400 people corresponding to the 10 Barcelona districts, a city division coarser than that of neighbourhoods. This design was followed in order to achieve given precision levels for the city districts. Our goal is now to take that sampling design into account for inference. In addition, although it was ignored when designing the sample, we have an additional auxiliary variable that could be taken into account for inference. This binary variable collects whether or not each individual owns their place of residence. The Barcelona health survey specifically asks this question in each survey and the city census contains the total of this same variable for each neighborhood of the city ($\{N_i^{home=k},\;i=1,\ldots,73,\;k=1,2\}$). Thus, this variable could be used to derive a ratio estimator that includes information on the ownership of the homes and its relationship with the birth country of Barcelona inhabitants. Additionally, a combined estimator could be also derived that takes into account the age stratification design of the sample, the home ownership information available and spatial dependence and integrates it all into a single model. This is the main goal of this case study. For simplicity, and due to the mild association found between sex and the LMIC origin in Barcelona residents, we will ignore the sex stratification of the sampling design throughout the following analysis.

We have derived and compared three estimators in this case study. First, we have developed stratified estimates of the proportion of LMIC people for each neighborhood $\boldsymbol{\hat{\pi}^{Str}}$ (see Expression (\ref{EqStr})), taking the age stratification design into account and ignoring both spatial dependence and home ownership information. Second, we have developed a ratio estimator $\boldsymbol{\hat{\pi}^{Ratio}}$ (see Expression (\ref{EqRatio})), ignoring both spatial dependence and age stratification, but considering the home ownership information available from both information sources (Barcelona health survey and register of inhabitants (for the totals of the variable)). Finally, we have developed a third estimator $\boldsymbol{\hat{\pi}^{Full}}$ considering age strata, home ownership information and spatial dependence. This estimator has been derived by fitting a Bayesian hierarchical model which assumes:
$$Y_{ij}^{age=k_1 \;home=k_2} \sim Bernoulli((\pi^*)_i^{age=k_1 \;home=k_2})$$
$$logit((\pi^*)_{i}^{age=k_1 \;home=k_2})=\mu+\psi_{i}+\phi_{k_1}+\varphi_{k_2},$$
for $i=1,\ldots,73$, $j=1,\ldots,n_i$, $k_1=1,\ldots,5$ and $k_2=1,2$. In this case, $\boldsymbol{\psi}$ models the spatial variability between neighborhoods,$\boldsymbol{\phi}$ the effect of the age group and $\boldsymbol{\varphi}$ the effect of home ownership. Once this model has been fitted we could add all these probabilities for the different levels of the auxiliary variables into a single estimate per neighborhood ($\boldsymbol{\hat{\pi}}^{Full}$), taking into account (or not) finite sample corrections, as described above (Expression (\ref{FullEstimate})). Specifically, given the limited impact of the finite sample correction in the previous case study, we have not considered that correction when deriving $\boldsymbol{\hat{\pi}}^{Full}$ in this new study. Additionally, we have used Expression (\ref{IndEstimate}), which assumed independence between both auxiliary variables for deriving $\boldsymbol{\hat{\pi}}^{Full}$ since the population register of Barcelona does not publish population data for the combination of these variables at the neighborhood level, just the separate frequencies (by neighborhood) for each of them. The comparison of the standardized, ratio and full estimates, among them and against the estimates derived in the previous section, will allow their particular advantages to be assessed. The \texttt{R} code used for this analysis is also supplied as annex supplementary material to the paper.

As an initial consequence of using auxiliary variables for inference in our case study, we find that some combinations of these variables with the neighborhood are not present in the sample. Therefore, the proportions of LMIC may not be estimated in those neighborhoods for the $\hat{\boldsymbol{\pi}}^{Str}$ and $\hat{\boldsymbol{\pi}}^{Ratio}$ estimators. Specifically, for the age-stratified estimator $\hat{\boldsymbol{\pi}}^{Str}$ we find that 6.6

Table \ref{tableRatios} contains the same statistics as those shown in Table \ref{tableSRS} but for the new models developed in this section. In addition, this table also contains two columns corresponding to the $\boldsymbol{\hat{\pi}^{SRS}}$ estimator of the previous section, but considering as missing values those positions where $\boldsymbol{\hat{\pi}^{Str}}$ (13 missing values) and $\boldsymbol{\hat{\pi}^{Ratio}}$ (4 missing values) are also missing, respectively. This allows the stratified and ratio estimators to be compared with those that considered sampling random sampling over the same set of spatial units.

\begin{table}
\centering
\caption{Assessment criteria for the stratified, ratio and full proportion estimates}
\begin{tabular}{ |c|c|c|c|c|c| }
 \hline
 Statistic & $\boldsymbol{\hat{\pi}^{SRS}}$  & $\boldsymbol{\hat{\pi}^{Str}}$ & $\boldsymbol{\hat{\pi}^{SRS}}$  & $\boldsymbol{\hat{\pi}^{Ratio}}$ & $\boldsymbol{\hat{\pi}^{Full}}$\\
  & (13 missing) & & (4 missing) & &\\
  \hline
 $Cor(\boldsymbol{\pi},\hat{\boldsymbol{\pi}})$ & 0.741 & 0.724 & 0.699 & 0.727 & 0.796\\
 $\sqrt{MSE}$ & 0.061 & 0.062 & 0.069 & 0.062 & 0.043\\
 95\% CI length & 0.206 & 0.101 & 0.232 & 0.115 & 0.141\\
 95\% CI Coverage & 0.867 & 0.600 & 0.826 & 0.667 & 0.959\\
 Moran's I & 0.048 & 0.044 & 0.158 & 0.150 & 0.305\\
 P(estimated I<real I) & 1 & 1 & 0.838 & 0.806 & 0.643\\
 \hline
\end{tabular}
\label{tableRatios}
\end{table}

The comparison of the second and third columns of Table \ref{tableRatios} shows that there is hardly any difference between the stratified estimator and SRS in terms of the correlation and MSE with respect to $\boldsymbol{\pi}$. Thus, although SRS estimates were in principle biased due to ignoring the stratified sample design followed, the effect of ignoring that design in $\boldsymbol{\hat{\pi}^{SRS}}$ does not seem to yield a high bias. In contrast, differences are more evident in terms of confidence intervals since the lengths of the stratified estimates are substantially shorter than those assuming SRS. Although seems like good news, this increased precision also reduces the proportion of times that the ICs contain the real values $\boldsymbol{\pi}$ from 0.867 to 0.600, far lower than expected. This could be a consequence of the small size of the units of study, which, under stratified sampling, means that many combinations of age groups and neighborhoods have really low sample sizes. Specifically, 28.5

Columns 4 and 5 of Table \ref{tableRatios} show a similar performance of the ratio estimator to that of $\boldsymbol{\hat{\pi}^{Str}}$, although the ratio estimator displays slight improvements. Thus, $\boldsymbol{\hat{\pi}^{Ratio}}$ estimates $\boldsymbol{\pi}$ similarly to $\boldsymbol{\hat{\pi}^{SRS}}$, though it is slightly better in terms of $Cor(\boldsymbol{\pi},\hat{\boldsymbol{\pi}})$ and MSE. Once again the CI lengths are reduced in comparison to SRS, but that reduction is lower (and, correspondingly, the CI coverages are higher) than for the stratified estimators. This could be explained by the 8.9

Finally, the 6th column of Table \ref{tableRatios} illustrates $\boldsymbol{\hat{\pi}^{Full}}$ results. This estimator is clearly better than $\boldsymbol{\hat{\pi}^{Str}}$ and $\boldsymbol{\hat{\pi}^{Ratio}}$ for all the criteria considered, except perhaps the CI lengths (which are wider), which, as shown, seems to be superior in terms of the CI coverage. According to the Moran's I statistic, $\boldsymbol{\hat{\pi}^{Full}}$ also appears to have a spatial dependence comparable to that of $\boldsymbol{\pi}$. Additionally, $\boldsymbol{\hat{\pi}^{Full}}$ also seems to display significant advantages in comparison to the SRS spatial estimator (Table \ref{tableSRS}) for all the criteria considered. Therefore, merging the advantages of stratified, ratio and small areas modeling into a single proposal appears to yield a clear reward for small area modeling, which is made possible by the use of Bayesian hierarchical models.

\begin{figure}
\centering
\includegraphics{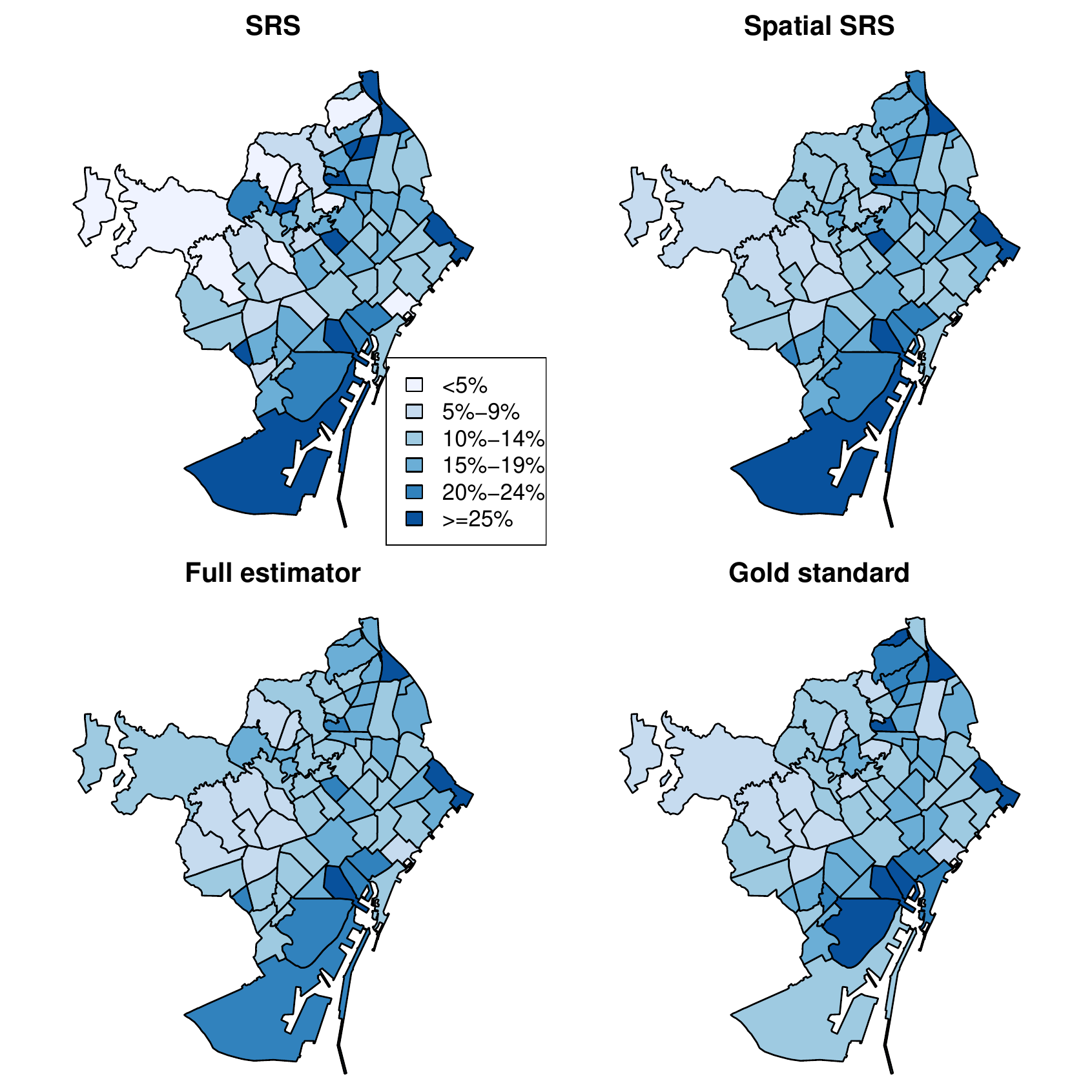}
\caption{Estimated spatial patterns for the percentage of low- and middle-income countries population for several of the models in this paper, and that same percentage according to the Barcelona census gold standard.}
\label{fig:FigMaps}
\end{figure}

Figure \ref{fig:FigMaps} shows several choropleth maps with some of the estimated proportions in this work, as well as with the census gold standard. This figure illustrates the evident noise of the SRS estimates when no specific small area methods are used to estimate the LMIC population proportions. This noise is filtered out for the spatial and full estimates where the proportion estimates show, in addition, a similar spatial dependence to that displayed by the gold standard geographical pattern. Differences between the spatial and full estimators are visually more subtle, indicating that the greatest benefit of the full estimator may come from the modeling of the spatial dependence. In any event, the main advantage of taking auxiliary variables into account could be the increase in the precision of the proportion estimates, which seems confirmed by the CI lengths of the spatial and full models in Tables \ref{tableSRS} and \ref{tableRatios}. These different precision estimates cannot be directly noticed in Figure \ref{fig:FigMaps}.

\section{Conclusions}

This paper proposes merging both survey sampling and spatial small area approaches in order to derive sensible estimates from a survey sample over a collection of small areas. Our approach describes how to include information from auxiliary variables for making inference on the outcome of interest, whether these were used for the sampling design (stratification variables) or not (ratio estimators). We also propose merging sampling theory with spatial methods in order to yield improved estimates which enjoy the advantages of both approaches. We propose combining the theoretical estimates from sampling theory corresponding to these two settings (stratified and ratio estimators) with spatial methods which yield dependent estimates of the underlying group-specific proportions for each small area. The use of spatial methods to avoid the independence assumption of those underlying probabilities is the key to deriving sensible estimates in survey-based small area studies since that independence assumption, widespread for classical sampling theory estimators, could clearly be harmful in the small area context. As shown, the potential bias introduced by spatial dependence has many more advantages than drawbacks when dealing with small areas. Nevertheless, beyond the spatial stratified and ratio estimators derived in this paper, our main goal in this work is to merge sampling theory estimation with spatial statistics methods into a single proposal. Following our proposal, similar approaches could be undertaken for different or more complex sampling designs in small area studies, such as for proportional to size sampling, regression estimators ... These settings would deserve similar attention to that paid in our paper to stratification or ratio estimators and, in principle, could be successfully modeled following similar procedures.

Typically, when a sampling design has been followed, the main goal of inference methods is to take that design into account in order to yield appropriate estimates. Nevertheless, the latest of our case studies shows that the use of auxiliary variables that were not used for sampling design (home ownership status in our case) could be even more advantageous in practice than the control of sampling design variables (age group). In addition, we have also shown that spatial dependence could also yield substantial benefits, even greater than the control of sampling design variables, since in our case the spatial estimates clearly outperform stratified and ratio estimates. Evidently, this would probably depend on each specific data set, in particular on whether the design variable is closely related, or not, to the outcome variable. However, a clear message seems to arise here: when dealing with small areas, the modeling of the outcome variable, in particular its dependence structure, could be as important as taking the sampling design into account. Therefore, sampling theory alone does not appear to be sufficient in and of itself to derive appropriate estimates in small areas studies. In those cases, modeling resources, such as including spatial dependence or auxiliary variables which were not used for the sampling design, are important tools that should be used whenever possible in order to yield sensible estimates.

In this paper we propose modeling the underlying probabilities for each area and group of the auxiliary variable(s) by considering them as spatially dependent, which in principle seems to solve small area estimation problems. Nevertheless, more elaborate spatial modeling could be used in order to yield more flexible estimates for these probabilities. Thus, in the previous section, we proposed modeling those specific group-area probabilities as $logit((\pi^*)_i^{Z=k})=\mu+\psi_i+\phi_k$, which cannot reproduce interactions between age groups and spatial units. We could alleviate this constraint by modeling $logit((\pi^*)_i^{Z=k})=\mu_i+\xi_{ik}$, where the rows of $\boldsymbol{\xi}$ are modeled by means of a spatiotemporal \citep{Martinez-Beneito.LopezQuilez.ea2008} or a multivariate CAR process \citep{Botella-Rocamora.Martinez-Beneito.ea2015}, which allows the spatial patterns for the different age groups to be different, though dependent. In a similar manner, if two (or more) auxiliary variables were used for inference, such as for the second of our case studies, multidimensional spatial modeling \citep{Martinez-Beneito.Botella-Rocamora.ea2017} could be used in order to model $(\pi^*)_{i}^{Z_1=k_1\;Z_2=k_2}$ as $logit((\pi^*)_{i}^{Z_1=k_1\;Z_2=k_2})=\mu+\xi_{ik_1k_2}$. In this expression, $\xi_{ik_1k_2}$ will be now a three-dimensional array, with a different, but dependent, spatial pattern for the different values of $k_1$ and $k_2$. Clearly, these models would yield more flexible estimates of the underlying group-area specific probabilities, which would make the area-specific estimates $\boldsymbol{\hat{\pi}^{Full}}$ less constrained, while taking advantage of spatial dependence, which has been shown to be so advantageous. That enhancement of our current proposal is beyond the scope of this paper and should possibly be explored in the future. In any event, the current proposal in this paper, although perhaps somewhat restrictive in comparison to the multivariate and multidimensional alternatives, has been proven to be interesting in and of itself, with quite promising results in practice. As shown, when dealing with small areas, spatial modeling, such as that introduced in this paper, seems to be the most advisable option, much more advisable than the use of traditional direct estimates supported by sampling theory.

\bibliographystyle{unsrt}
\bibliography{BibliografiaUTF8.bib}

\begin{thebibliography}{10}

\bibitem{Martinez-Beneito.BotellaRocamora2019}
Miguel~A. Martinez-Beneito and P.~Botella~Rocamora.
\newblock {\em Disease mapping from foundations to multidimensional modeling}.
\newblock CRC Press, 2019.

\bibitem{Lawson2018}
Andrew~B. Lawson.
\newblock {\em Bayesian Disease Mapping: Hierarchical Modeling in Spatial
  Epidemiology (3rd edition)}.
\newblock CRC Press, 2018.

\bibitem{Hansen.Hurwitz1943}
Morris~H. Hansen and William~N. Hurwitz.
\newblock On the theory of sampling from finite populations.
\newblock {\em The Annals of Mathematical Statistics}, 14(4):333--362, 1943.

\bibitem{Horvitz.Thompson1952}
D.~G. Horvitz and D.~J. Thompson.
\newblock A generalization of sampling without replacement from a finite
  universe.
\newblock 47(260):663--685.

\bibitem{Swensson.Saerndal.ea2003}
Bengt Swensson, Carl-Erik Särndal, and Jan Wretman.
\newblock {\em Model Assisted Survey Sampling}.
\newblock Springer New York, 2003.

\bibitem{Chen.Wakefield.ea2014}
Cici Chen, Jon Wakefield, and Thomas Lumely.
\newblock The use of sampling weights in bayesian hierarchical models for small
  area estimation.
\newblock 11:33--43.

\bibitem{Mercer.Lu.ea2019}
Laina~D. Mercer, Fred Lu, and Joshua~L. Proctor.
\newblock Sub-national levels and trends in contraceptive prevalence, unmet
  need, and demand for family planning in nigeria with survey uncertainty.
\newblock 19(1).

\bibitem{Paige.Fuglstad.ea2020}
John Paige, Geir-Arne Fuglstad, Andrea Riebler, and Jon Wakefield.
\newblock Design- and model-based approaches to small-area estimation in a low-
  and middle-income country context: Comparisons and recommendations.

\bibitem{Little1993}
R.~J.~A. Little.
\newblock Post-stratification: A modeler{\textquotesingle}s perspective.
\newblock 88(423):1001--1012.

\bibitem{Park.Gelman.ea2004}
David~K. Park, Andrew Gelman, and Joseph Bafumi.
\newblock Bayesian multilevel estimation with poststratification: State-level
  estimates from national polls.
\newblock 12(4):375--385.

\bibitem{Gelman2007}
Andrew Gelman.
\newblock Struggles with survey weighting and regression modeling.
\newblock 22(2).

\bibitem{Lohr2010}
Sharon~L. Lohr.
\newblock {\em Sampling: design and analysis (2nd edition)}.
\newblock Brooks/Cole, Cengage Learning, 2010.

\bibitem{Kunsch1987}
Hans~R Kunsch.
\newblock Intrinsic autoregressions and related models on the two-dimensional
  lattice.
\newblock {\em Biometrika}, 74:517--524, 1987.

\bibitem{Leroux.Lei.ea1999}
Brian~G. Leroux, Xingye Lei, and Norman Breslow.
\newblock Estimation of disease rates in small areas: a new mixed model for
  spatial dependence.
\newblock In M~E Halloran and D~Berry, editors, {\em Statistical Models in
  Epidemiology, the Environment and Clinical Trials}. Springer, Berlin
  Heidelberg New York, 1999.

\bibitem{Besag.York.ea1991}
Julian Besag, Jeremy York, and Annie Molli\'{e}.
\newblock Bayesian image restoration, with two applications in spatial
  statistics.
\newblock {\em Annals of the Institute of Statistical Mathemathics}, 43:1--21,
  1991.

\bibitem{Moran1948}
P.~A.~P. Moran.
\newblock The interpretation of statistical maps.
\newblock 10:243--251.

\bibitem{MacNab.2018}
Ying~C. MacNab.
\newblock Some recent work on multivariate {G}aussian {M}arkov random fields.
\newblock {\em TEST}, 27(3):497--541, 2018.

\bibitem{Martinez-Beneito.LopezQuilez.ea2008}
Miguel~A. Martinez-Beneito, Antonio L\'{o}pez-Qu\'{i}lez, and Paloma
  Botella-Rocamora.
\newblock An autoregressive approach to spatio-temporal disease mapping.
\newblock {\em Statistics in Medicine}, 27:2874--2889, 2008.

\bibitem{Botella-Rocamora.Martinez-Beneito.ea2015}
Paloma Botella-Rocamora, Miguel~A. Martinez-Beneito, and Sudipto Banerjee.
\newblock A unifying modeling framework for highly multivariate disease
  mapping.
\newblock {\em Statistics in Medicine}, 34(9):1548--1559, 2015.

\bibitem{Martinez-Beneito.Botella-Rocamora.ea2017}
Miguel~A Martinez-Beneito, Paloma Botella-Rocamora, and Sudipto Banerjee.
\newblock Towards a multidimensional approach to {B}ayesian disease mapping.
\newblock {\em Bayesian Analysis}, 12:239--259, 2017.

\end{thebibliography}

\end{document}